\documentclass[reprint, aps, prb, superscriptaddress, twocolumn, showpacs]{revtex4-1}

\usepackage{natbib}
\usepackage{graphicx}
\usepackage[ansinew]{inputenc}
\usepackage{bm}
\usepackage{epstopdf}

\newcommand{\eg}{{\em, e.\,g.,\ }}
\newcommand{\unit}[1]{\text {\,#1}}
\newcommand{\etal}{{\em et al.~}}

\newcommand{\Gammabar}{\ensuremath{\overline{\Gamma}}}

\newcommand{\E}[2]{\ensuremath{#1 \cdot 10^{#2}}}
\newcommand{\didv}{\ensuremath{\text{d}I/\text{d}V}}
\newcommand{\RM}[1]{\MakeUppercase{\romannumeral #1}}
\newcommand{\ind}[1]{\ensuremath{_{\textnormal{\scriptsize{#1}}}}}

\begin{document}

\title{
Scanning Tunneling Spectroscopy of Ni/W(110):\\bcc and fcc properties in the second atomic layer
}

\author{Johannes Schöneberg}
\email{schoeneberg@physik.uni-kiel.de}
\author{Alexander Weismann}
\author{Richard Berndt}
\affiliation{Institut für Experimentelle und Angewandte Physik, 
Christian-Albrechts-Universität zu Kiel, D-24098 Kiel, Germany}

\begin{abstract}
Nickel islands are grown on W(110) at elevated temperatures.
Islands with a thickness of two layers are investigated with scanning tunneling microscopy.
Spectroscopic measurements reveal that na\-no\-me\-ter si\-zed areas of the islands exhibit distinctly different apparent heights and \didv spectra. Spin polarized and paramagnetic band structure calculations indicate that the spectral features are due to fcc(111) and bcc(110) orientations of the Ni film, respectively.
\end{abstract}

\maketitle
\section{Introduction}
\label{intro}

The magnetism of thin iron, nickel and cobalt films differs strongly from the behavior of the bulk materials  \cite{RPP.71.056501,PRL.73.898,PRB.49.3962}.
Quantities like the Curie temperature \cite{PRL.68.1208} and the magnetic anisotropies \cite{PRB.51.15933} show thickness dependent characteristics.
Since the real space structure and the magnetic properties are connected, most of these effects can be explained by the atomic structure. Different lattice constants of the substrate and film materials can result in different crystallographic orientations and cause strain, which is a source of magnetic anisotropy.
The strain is released when a transition occurs from pseudomorpic growth to a bulk-like film.
This has been intensively studied using $k$-space methods\eg low energy electron diffraction (LEED) \cite{SS.144.495} and angle resolved photoelectron spectroscopy (ARPES) \cite{PRB.38.9451}. 

Even within the same layer different crystallographic orientations may coexist. This was observed for Fe on Cu(111) \cite{NN.5.792} (bcc and fcc), Fe on Ir(111) \cite{NJP.9.396} and Co on W(110) \cite{PRB.72.035460} (both fcc and hcp) using scanning tunneling microscopy (STM).
These studies discriminated both phases by observing different island orientations, by using atomically resolved topographies, or by resolving different electronic structures using scanning tunneling spectroscopy (STS).

A common substrate for growing films of 3d metals is W which shows almost no interdiffusion.
Due to its high surface energy, adsorbates tend to form a wetting layer.
For Fe the growth of subsequent layers depends highly on the temperature \cite{PRB.54.R8385,JVSTA.15.1285}.

The growth of Ni on W(110) at room temperature has been studied with the STM by Sander, Schmidthals \etal \cite{PRB.57.1406,SS.402.636,SS.417.361}.
Here we present STM data of Ni grown at elevated temperatures.
This investigation was motivated by the possibility to obtain bcc Ni similar to the case of Co on W(110) which forms a pseudomorphic and a close packed crystallographic structure in the sub-monolayer regime \cite{PRB.67.153405}.
Bcc Ni has attracted much attention for an expected phase transition from paramagnetic to ferromagnetic depending on the lattice constant \cite{PRL.57.2211,PRB.34.1784,PRB.38.1613}.
Investigations of bcc Ni on Fe \cite{JVSTA.4.1376,SSS.61.623,PRB.46.237} and GaAs \cite{PRL.94.137210} substrates addressed this issue with spatially averaging techniques.
To obtain this Ni phase we focused on the second atomic layer (AL), which is the threshold thickness for Ni growing in a relaxed fcc fashion with only small tensile strain as indicated by low-energy electron diffraction and measurements of the film stress \cite{SS.417.361}.

Using magnetic as well as non-magnetic tips and comparing our results to bandstructure calculations we find that islands of the second monolayer are comprised of areas of fcc and bcc Ni.

\section{Experimental Setup}
\label{setup}

The experiments were performed using a home-built STM operating at $4.4\unit{K}$ in ultrahigh vacuum. 
W(110) surfaces were cleaned by heating to a temperature of $1500\unit{K}$ in an O$_{2}$ atmosphere with a partial pressure of $\E{1}{-4}\unit{Pa}$ followed by short heating to $2300\unit{K}$.
The absence of contamination was checked with LEED and Auger electron spectroscopy.
Nickel films were deposited at temperatures of $500\unit{K}$ using an electron beam evaporator and an evaporant of 99.99\% purity. Electrochemically etched tungsten and nickel tips were prepared \emph{in situ} by Ar$^{+}$ bombardement and annealing.
Ni tips were magnetized along the tip axis with a CoSm magnet.

\section{Results}
\label{results}

\begin{figure}
	\centering
		\includegraphics[width=85mm]{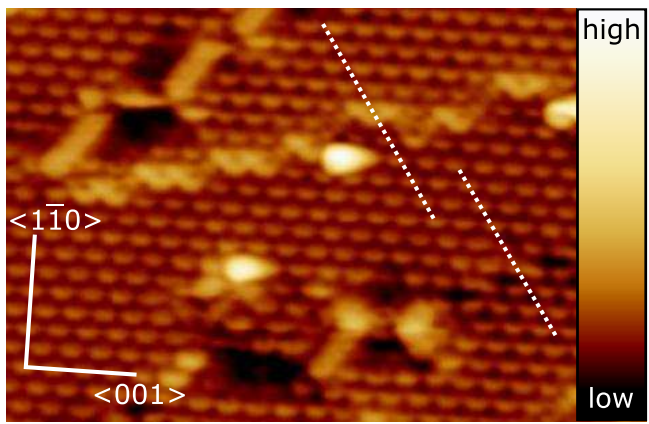}
	\caption{Constant-current STM image of the wetting layer Ni on W(110)
	($12.5 \times 9\unit{nm}^2, 100\unit{mV}, 50\unit{pA}$).
	A $2\times2$ reconstruction is resolved along with a few antiphase domain boundaries.
	The dashed lines are used to indicate the lattice offset occurring across boundaries.}
	\label{fig:Fig1}
\end{figure}

Figure\ \ref{fig:Fig1} shows a constant current topograph of the wetting layer.
In contrast to the results published for films grown at room-temperature \cite{PRB.57.1406}, the Ni wetting layer forms a $2\times2$ reconstruction when deposited at $500\unit{K}$.
The interatomic distances in the $\left\langle1\overline{1}0\right\rangle$ and $\left\langle001\right\rangle$ directions only deviate by 5\% and 1\% from $2\sqrt{2}a\ind{W}$ and $2a\ind{W}$, respectively ($a\ind{W}$: lattice constant of the conventional unit cell of W). 
Low energy electron diffraction (LEED) patterns (not shown) confirmed this surface structure.
Along the $\left\langle1\overline{1}0\right\rangle$ directions the reconstruction is perturbed by antiphase boundaries which occur in distances of few nm (dashed lines in Fig.\ \ref{fig:Fig1}(a)).
Since a $2\times2$ reconstruction covers the surface only by a quarter we use the terminology of Ref.\ \onlinecite{SS.417.361} and refer to observed layers as \emph{atomic layers}. 
By the term \emph{monolayer} the amount of atoms for a fully closed layer with the atomic density of bulk Ni(111) ($\E{1.86}{19} \unit{atoms/}\unit{m}^{2}$) is meant. 
In several preparations we evaporated different coverages ranging from $\approx0.25$ to $\approx0.35$ monolayers.

\begin{figure}
	\centering
	\includegraphics[width=85mm]{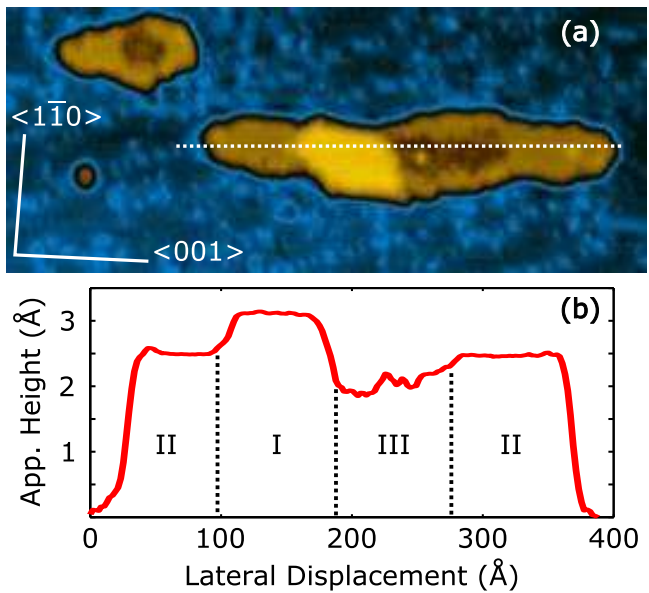}
	\caption{(a) Constant-current STM image of two Ni islands of 2 AL thickness on W(110) ($54 \times 22\unit{nm}^2$, $200\unit{mV}$, $1\unit{nA}$).
	The larger, elongated island exhibits four areas with three different apparent heights.
	(b) Linescan along the dashed line shown in (a).}
	\label{fig:Fig2}
\end{figure}

Fig.\ \ref{fig:Fig2}(a) shows a constant-current topograph of two islands, which are characteristic of the second atomic layer.
The islands are mostly oriented along the W$\left\langle001\right\rangle$ directions in agreement with Ref.\ \onlinecite{SS.417.361}. In contrast to this study we observe a detailed structure:  The larger island in Fig.\ \ref{fig:Fig2} (a) exhibits four areas with three different apparent heights. The areas are labeled \RM{1}, \RM{2}, and \RM{3} in the cross-section (Fig.\ \ref{fig:Fig2}(b)).
Area \RM{2} appears $\approx 2.5\unit{\AA}$ higher than the Ni wetting layer.
The areas \RM{1} and \RM{3} are $\approx 0.5 \unit{\AA}$ higher and lower, respectively, than area \RM{2}.
The height difference is voltage dependent:  For voltages $-1\unit{V} \dots 1\unit{V}$ area \RM{1} is $0.25\unit{\AA}$ to $0.8\unit{\AA}$ higher than area \RM{3}. The maximum difference of $0.8\unit{\AA}$ is measured at a voltage of $200\unit{meV}$.
Given the interlayer distance for Ni fcc(111) of $\approx2\unit{\AA}$ the height variations cannot be attributed to by different layer thicknesses.
Therefore we hint that similar to Co/W(110) \cite{PRB.67.153405} different crystallographic structures are present.

\begin{figure*}
	\centering
		\includegraphics[width=170mm]{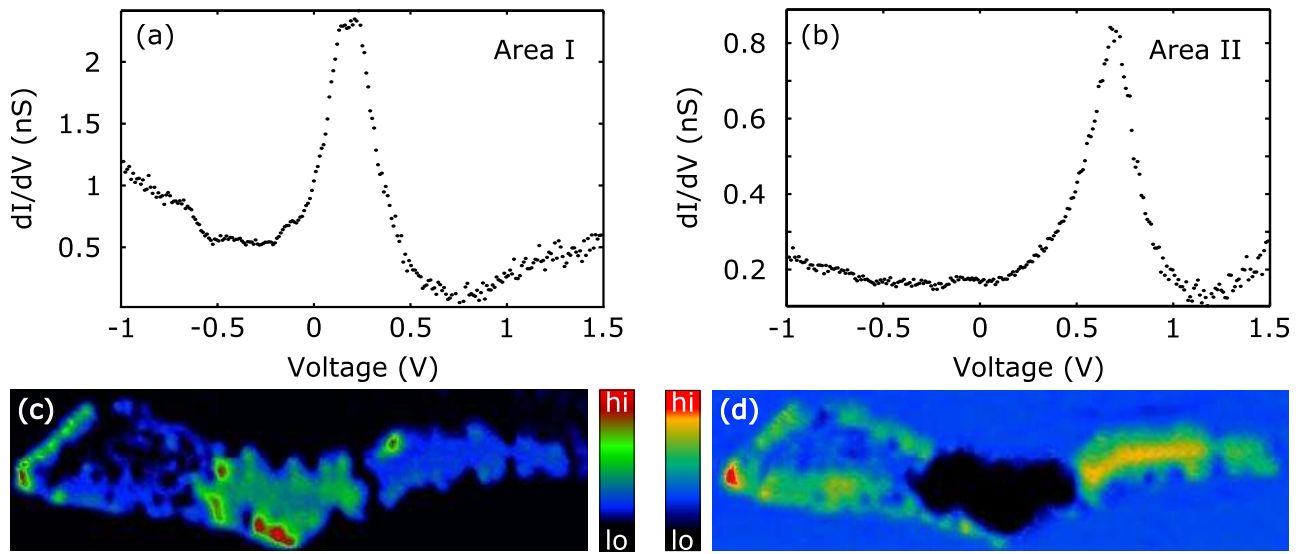}
	\caption{(a) and (b):
	\didv spectra of a second AL island recorded on areas \RM{1} and \RM{2}, respectively.
	(parameters prior to opening the current feedback loop: $1.5\unit{V}$, $1\unit{nA}$)
	(c) and (d): corresponding \didv maps $42 \times 12\unit{nm}^2$, recorded at $200\unit{mV}$ and $700\unit{mV}$, respectively. The island consists of area \RM{3}, \RM{1} and \RM{2} from left to right.}
	\label{fig:Fig3}
\end{figure*}

Areas \RM{1} and \RM{2} exhibit smooth surfaces. Further investigations using STS show prominent peaks in the differential conductances (\didv, $I$ current, $V$ sample voltage) (Figs.\ \ref{fig:Fig3}(a, b)).

The peaks occur at $200\unit{mV}$ and $700\unit{mV}$ in areas \RM{1} and \RM{2}, respectively, which explains the maximum of the apparent height at $200\unit{mV}$.
They were observed with spin-polarized Ni tips as well as with spin-averaging W tips.
The signals are therefore not related to different local spin polarizations. 
This interpretation is further supported by the large energy difference of $500\unit{meV}$ between both peaks, which drastically exceeds the $\approx 10 \unit{mV}$ shift reported for domains in two AL of Fe on W(110) \cite{APL.96.132505}.
\didv~ maps recorded at 200 and $700\unit{mV}$ (Figs.\ \ref{fig:Fig3}(c, d)) reveal that these unoccupied states are characteristic of the different areas of the island.
They also show that the edge of area \RM{3} has the similar characteristics as area \RM{2}, which was further confirmed by taking \didv~spectra directly above the edge of \RM{3}.
In high resolution STM images area \RM{3} appears rather disordered with a standard deviation of the apparent height of $\approx0.14\unit{\AA}$. Thus this area presently cannot be attributed to a particular crystallographic surface.
Spectra of the differential conductance (not shown) are featureless in the range $-1.5\unit{V}$ to $1.5\unit{V}$. 
We hint that this area is disordered due to underlying antiphase boundaries.

Ref.\ \onlinecite{SS.417.361} suggested that the second atomic layer of Ni grows in a fcc phase, while we observe different structures at this coverage.
A possible reason for this discrepancy may be the different techniques used in identifying the structure. We locally probed Ni islands grown at elevated temperatures while the analysis in Ref.\ \onlinecite{SS.417.361} is based on spatially averaging methods (LEED, stress-measurements).

\begin{figure}
	\centering
		\includegraphics[width=85mm]{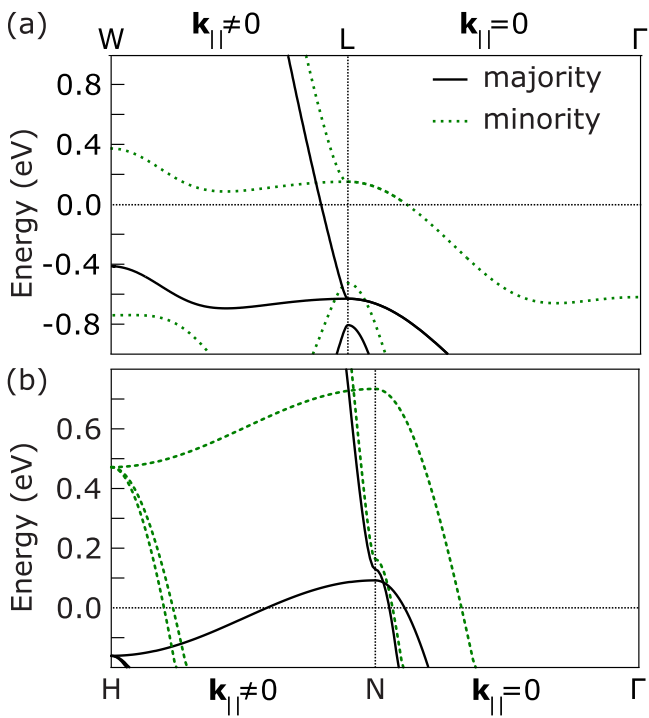}
	\caption
	{
	Calculated spin-polarized bulk band structure of Ni.
	(a) LCAO calculation for the fcc configuration. 
		Algorithm and tight-binding parameters from Ref.\ \onlinecite{Book.Papaconst}.	
	(b)DFT LSDA calculation for the bcc configuration reproduced from Ref.\ \onlinecite{PhD.Ohm} 
	}
	\label{fig:Fig4}
\end{figure}

To link areas \RM{1} and \RM{2} to crystallographic surfaces we calculated the spin-polarized bulk band structure.
Fig.\ \ref{fig:Fig4}(a) shows results from a LCAO (linear combination of atomic orbitals) calculation of fcc Ni using a three-center-approximation with non-orthogonal basis set.

The algorithm and the values for the tight-binding parameters, matching \emph{ab-initio} calculations, were taken from Ref.\ \onlinecite{Book.Papaconst}. This band structuture is further validated by photoemmission studies \cite{PRB.19.2919} and other calculations \cite{PRB.15.298}.

The tunneling current is predominantly due to states at the center of the surface Brillouin zone (\Gammabar).
At a fcc(111) surface these are the states between the high symmetry points $\Gamma$ and $L$ of the bulk Brillouin zone, which are projected onto the \Gammabar~ point.
A minority spin band exhibits a band edge above the Fermi energy at $\approx150 \unit{meV}$, which is close to the experimental \didv peak at $200\unit{mV}$ (Fig.\ \ref{fig:Fig3}(a)). Also note that the band edges at $\approx-500\unit{meV}$ correspond to the enhanced \didv signal in Fig.\ \ref{fig:Fig3}(a) at $-600\unit{mV}$.
Area \RM{1} is therefore attributed to a fcc Ni(111) surface.
The approximation of using bulk band structures is supported by previous stress measurements suggesting that Ni grows in a relaxed fashion from the second atomic layer on with only small tensile film stress \cite{SS.417.361}.
Moreover, Ni is known to grow in a fcc(111) orientation for higher coverages \cite{SS.417.361}. Furthermore, according to ARPES studies \cite{PRB.38.9451}, the density of states of the fifth atomic layer Ni on W(110) and bulk Ni are almost indistinguishable.
We therefore prepared thicker films with local coverages up to 8 atomic layers.
In full agreement with the data from area \RM{1} of second atomic layer islands the \didv measurements on thicker films exclusively show a peak at $200\unit{mV}$. A peak at $700\unit{mV}$ was not observed.
These data show that the unoccupied state at $200\unit{meV}$ is characteristic of fcc Ni(111). This further validates the identification of area \RM{1} with fcc Ni(111) and the band structure calculations we employed seem to be valid in the limit of the second atomic layer. 

Since the band structure of the fcc crystal shows no band edges which would correspond to the peak at $700\unit{meV}$ we suggest that area \RM{2} has no fcc structure.
A tungsten induced bcc-like structure similar to the pseudomorphic grown layer reported for Co/W(110) \cite{PRB.67.153405} may be present. Furthermore bcc Ni was grown successfully on Fe(001) \cite{JVSTA.4.1376,SSS.61.623,PRB.46.237} and GaAs(001) \cite{PRL.94.137210}.
In order to gain more insight Fig.\ \ref{fig:Fig4}(b) reproduces results from a density functional theory (DFT) calculation for bcc Ni using the local spin density approximation (LSDA) from Ref.\ \onlinecite{PhD.Ohm}.
Due to the orientation of the W crystal a bcc(110) surface is the most probable case. Here the states between the high symmetry points $\Gamma$ and $N$ of the bulk Brillouin zone are projected onto the \Gammabar~ point.
A minority spin band exhibits an edge at $\approx730\unit{meV}$.
The corresponding high density of states is consistent with the \didv\ peak observed at $700\unit{mV}$ (Fig.\ \ref{fig:Fig3}(b)).
However the majority band edge at $100\unit{meV}$ is not observed experimentally with both Ni and W tips.
Moruzzi \cite{PRL.57.2211,PRB.34.1784,PRB.38.1613} calculated that bcc Ni exhibits a first order transition between paramagnetic and ferromagnetic order with increasing lattice constant. Bcc Ni at equilibrium is predicted to be paramagnetic. In this case no spin splitting is present and the band edge is located at $\approx320\unit{meV}$ \cite{PhD.Ohm}.
In general there is a clear trend that the relevant band edges of bcc Ni occur at higher energies than for fcc Ni, independent whether the calculations were performed for the paramagnetic or the ferromagnetic case. Therefore we tentatively suggest that area \RM{2} is bcc Ni in a (110) orientation.

\section{Conclusions}
\label{conclusions}
Three different structures were observed in second atomic layer Ni islands, which were grown on W(110) at elevated temperatures.
This refines previous studies \cite{PRB.57.1406,SS.402.636,SS.417.361} where space-averaging methods resolved only a fcc phase in the second atomic layer.
While the lowest area shows a disordered structure with local apparent height variations of $\approx0.14\unit{\AA}$ and no features in \didv spectra, the other areas appear topographically smooth and exhibit prominent spectroscopic signatures.
By comparision with spin polarized band structure calculations the highest area is attributed to fcc Ni(111).
Calculations performed for Ni bcc indicate that the area showing a peak at $700\unit{meV}$ may be Ni bcc (110).

\section*{Acknowledgments}
Funding through SFB 668 of the Deut\-sche For\-schungs\-ge\-mein\-schaft is gratefully acknowledged.


\begin{thebibliography}{10}
\providecommand{\url}[1]{{#1}}
\providecommand{\urlprefix}{URL }
\expandafter\ifx\csname urlstyle\endcsname\relax
  \providecommand{\doi}[1]{DOI \discretionary{}{}{}#1}\else
  \providecommand{\doi}{DOI \discretionary{}{}{}\begingroup
  \urlstyle{rm}\Url}\fi

\bibitem{RPP.71.056501}
C.A.F. Vaz, J.A.C. Bland, G.~Lauhoff, Rep. Prog. Phys. \textbf{71}, 056501
  (2008)

\bibitem{PRL.73.898}
H.J. Elmers, J.~Hauschild, H.~H\"{o}che, U.~Gradmann, H.~Bethge, D.~Heuer,
  U.~K\"{o}hler, Phys. Rev. Lett. \textbf{73}, 898 (1994)

\bibitem{PRB.49.3962}
F.~Huang, M.T. Kief, G.J. Mankey, R.F. Willis, Phys. Rev. B \textbf{49}, 3962
  (1994)

\bibitem{PRL.68.1208}
Y.~Li, K.~Baberschke, Phys. Rev. Lett. \textbf{68}, 1208 (1992)

\bibitem{PRB.51.15933}
H.~Fritzsche, J.~Kohlhepp, U.~Gradmann, Phys. Rev. B \textbf{51}, 15933 (1995)

\bibitem{SS.144.495}
J.~Ko\l{}aczkiewicz, E.~Bauer, Surf. Sci. \textbf{144}, 495 (1984)

\bibitem{PRB.38.9451}
K.P. K\"{a}mper, W.~Schmitt, G.~G\"{u}ntherodt, H.~Kuhlenbeck, Phys. Rev. B
  \textbf{38}, 9451 (1988)

\bibitem{NN.5.792}
L.~Gerhard, T.K. Yamada, T.~Balashov, A.F. Tak\'{a}cs, R.J.H. Wesselink,
  M.~D\"{a}ne, M.~Fechner, S.~Ostanin, A.~Ernst, I.~Mertig, W.~Wulfhekel, Nat.
  Nanotechnol. \textbf{5}, 792 (2010)

\bibitem{NJP.9.396}
K.~von Bergmann, S.~Heinze, M.~Bode, G.~Bihlmayer, S.~Blügel, R.~Wiesendanger,
  New J. Phys. \textbf{9}, 396 (2007)

\bibitem{PRB.72.035460}
M.~Pratzer, H.J. Elmers, Phys. Rev. B \textbf{72}, 035460 (2005)

\bibitem{PRB.54.R8385}
M.~Bode, R.~Pascal, M.~Dreyer, R.~Wiesendanger, Phys. Rev. B \textbf{54}, R8385
  (1996)

\bibitem{JVSTA.15.1285}
M.~Bode, R.~Pascal, R.~Wiesendanger, J. Vac. Sci. Technol. A \textbf{15}, 1285
  (1997)

\bibitem{PRB.57.1406}
D.~Sander, C.~Schmidthals, A.~Enders, J.~Kirschner, Phys. Rev. B \textbf{57},
  1406 (1998)

\bibitem{SS.402.636}
C.~Schmidthals, A.~Enders, D.~Sander, J.~Kirschner, Surf. Sci \textbf{402}, 636
  (1998)

\bibitem{SS.417.361}
C.~Schmidthals, D.~Sander, A.~Enders, J.~Kirschner, Surf. Sci. \textbf{417},
  361 (1998)

\bibitem{PRB.67.153405}
M.~Pratzer, H.J. Elmers, Phys. Rev. B \textbf{67}, 153405 (2003)

\bibitem{PRL.57.2211}
V.L. Moruzzi, Phys. Rev. Lett. \textbf{57}, 2211 (1986)

\bibitem{PRB.34.1784}
V.L. Moruzzi, P.M. Marcus, K.~Schwarz, P.~Mohn, Phys. Rev. B \textbf{34}, 1784
  (1986)

\bibitem{PRB.38.1613}
V.L. Moruzzi, P.M. Marcus, Phys. Rev. B \textbf{38}, 1613 (1988)

\bibitem{JVSTA.4.1376}
B.~Heinrich, A.S. Arrott, J.F. Cochran, C.~Liu, K.~Myrtle, J. Vac. Sci.
  Technol. A \textbf{4}, 1376 (1986)

\bibitem{SSS.61.623}
Z.Q. Wang, Y.S. Li, F.~Jona, P.M. Marcus, Solid State Commun. \textbf{61}, 623
  (1987)

\bibitem{PRB.46.237}
N.B. Brookes, A.~Clarke, P.D. Johnson, Phys. Rev. B \textbf{46}, 237 (1992)

\bibitem{PRL.94.137210}
C.S. Tian, D.~Qian, D.~Wu, R.H. He, Y.Z. Wu, W.X. Tang, L.F. Yin, Y.S. Shi,
  G.S. Dong, X.F. Jin, X.M. Jiang, F.Q. Liu, H.J. Qian, K.~Sun, L.M. Wang,
  G.~Rossi, Z.Q. Qiu, J.~Shi, Phys. Rev. Lett. \textbf{94}, 137210 (2005)

\bibitem{APL.96.132505}
M.~Ziegler, N.~Ruppelt, N.~N\'{e}el, J.~Kr\"{o}ger, R.~Berndt, Appl. Phys.
  Lett. \textbf{96}, 132505 (2010)

\bibitem{Book.Papaconst}
D.A. Papaconstantopoulos, \emph{Handbook of the band structure of elemental
  solids} (Plenum Press, 1986)

\bibitem{PhD.Ohm}
T.~Ohm, Dft-Gutzwiller-Rechnungen zu ferromagnetischem bcc Nickel unter
  Ber\"{u}cksichtigung der Spin-Bahn-Kopplung.
\newblock Ph.D. thesis, University of Dortmund, Germany (2006)

\bibitem{PRB.19.2919}
F.J. Himpsel, J.A. Knapp, D.E. Eastman, Phys. Rev. B \textbf{19}, 2919 (1979)

\bibitem{PRB.15.298}
C.S. Wang, J.~Callaway, Phys. Rev. B \textbf{15}, 298 (1977)

\end{thebibliography}

\end{document}